\UseRawInputEncoding
\documentclass[%
 reprint,
 amsmath,amssymb,
 aps, physrev,
 prl
]{revtex4-2}

\usepackage{graphicx}
\usepackage{dcolumn}
\usepackage{bm}

\begin{document}

\preprint{APS/123-QED}

\title{\textbf{Regime identification and control of extremes in the non-autonomous Lorenz model with chaos and intransitivity} 
}%

\author{Moyan Liu$^{1}$}
\author{Qin Huang$^{1}$}
\author{Upmanu Lall$^{1,2}$}%
 \email{Contact author: ulall@asu.edu}
\affiliation{$^{1}$School of Complex Adaptive Systems \& Water Institute, Arizona State University, Tempe, AZ, USA}
\affiliation{$^{2}$Department of Earth and Environmental Engineering \& Columbia Water Center, Columbia University, New York, NY, USA}

\date{\today}

\begin{abstract}
Adaptive chaos control has been studied extensively for autonomous systems. For real world, non-autonomous systems, such as the planetary weather,  observations of the system state in response to seasonally and diurnally varying forcing are available only at discrete times and locations, over which system trajectories are likely to have diverged given uncertainties in initial conditions. We consider a stochastic representation of such systems, as a building block for adaptive control, and develop and test control strategies in an idealized setting. We present the first example of finite time adaptive chaos control for a seasonally forced and noise-perturbed Lorenz–84 model. We demonstrate two strategies for triggering control: (1) local Lyapunov exponents (LLE), and (2) transition probabilities for the latent states of a non-homogeneous Hidden Markov Model (NHMM). The second approach is motivated by thinking of future applications to a latent embedding space of planetary atmospheric circulation that would get us closer to real world analyses. The NHMM “triggers” are found to coincide with strongly positive LLE regimes, confirming their dynamical interpretability. These results provide a conceptual bridge towards the use of deep learning based weather and climate foundation models, whose hidden states could be leveraged for adaptive control to mitigate extreme weather events.
\end{abstract}

\maketitle


\section{\label{sec:level1}Introduction}
Over the past decades, research on chaos control has focused predominantly on autonomous systems \cite{BOCCALETTI2000103}. Researchers have developed chaos control methods such as the Ott–Grebogi–Yorke (OGY) method \cite{PhysRevLett.64.1196}, time-delayed feedback \cite{Purewal2016}, and model predictive control \cite{YU01041996}, demonstrating the feasibility of stabilizing chaotic systems through bounded interventions. More recently, Control Simulation Experiments (CSEs), ensemble-based predictive control, and data assimilation frameworks have been used to steer trajectory toward desired regimes \cite{npg-29-133-2022, npg-31-319-2024}. Applied to the Lorenz-63 (L63) model \cite{DeterministicNonperiodicFlow}, these methods have shown that regime switching can be suppressed, trajectories confined, and instabilities dampened in different aspects. 

However, many real-world systems are inherently non-autonomous, meaning that their dynamics are influenced by time-dependent forcings. Although non-autonomous systems have received attention \cite{5671133, Soong2007}, most control applications have been developed for autonomous chaotic systems. Weather dynamics are nonlinear with multiscale interactions that are modulated by seasonal and diurnal radiation forcing, leading to chaos, path dependence and intransitivity \cite{https://doi.org/10.1111/j.1600-0870.1984.tb00230.x,Lorezn1990}. How should chaos control be applied in this setting? 

We focus on this question in the context of real-world systems, where the dynamics are sampled at a finite frequency, which is typically much larger than the numerical integration time step used in model simulations. As a result, even between two consecutive observations, depending on the local Lyapunov exponent (LLE), the model simulated trajectories are expected to have diverged given uncertain initial conditions. Consequently, for effective control one needs to consider “noise” in the model to accommodate this divergence in the observational process, and/or to consider a stochastic control strategy. Previously, we introduced strategies \cite{egusphere-2025-3997} that employed the local Lyapunov exponent (LLE) as an instability indicator to trigger selectively optimized interventions in a noise forced version of autonomous Lorenz models. We extend this application to the non-autonomous version of the Lorenz-1984 (L84) model, and then consider a new approach that explicitly considers the stochastic observational process for adaptive control. 

The Non-homogeneous Hidden Markov Model (NHMM) \cite{https://doi.org/10.1111/1467-9876.00136, Kwon2009, Khalil23042010} is used as a general framework to represent these dynamics using a vector latent space that is identified directly from the observations. This model serves as a building block for future applications to higher dimensional latent states from deep learning models (e.g., Aurora, Pangu, AIFS) applied to spatio-temporal weather processes that currently require significant computational resources \cite{Bodnar2025,Bi2023,lang2024aifsecmwfsdatadriven}. In the present study, we assume that all state variables are directly observable, and therefore do not require attractor reconstruction or time-delay embedding from a single variable. However, the approach can be generalized to cases where only partial or single-variable observations are available. The NHMM encodes a low dimensional latent space and a Markovian transition probability across states that can vary over time, thus capturing the role of seasonality (non-autonomous factors) in the underlying dynamics. 

Climate extremes such as atmospheric rivers, hurricanes, heat waves, and freezes are intensifying in frequency and severity, producing devastating socio-economic impacts worldwide \cite{Pradhan2022, MARIO2024169269, Grant2025}. In this context, extreme events can be interpreted as corresponding to one of the latent states that is considered “dangerous”. The goal of the control strategy becomes the minimization of the probability of transitioning to such a state. The control is implemented using a trigger related to the probability of transition to a “dangerous” state, and small perturbations to the physical state variables are applied to nudge the system dynamics to other states. The objective of the control system is to minimize the total energy used in such nudges. We draw insights from an application of this approach and of an approach using LLE triggers, and find that the triggering conditions have some correspondence in either approach.

\section{\label{sec:level1}Methods}
We use the Lorenz-84 (L84) model of extratropical atmospheric circulation as a vehicle to develop the ideas. This model is introduced next, followed by a description of the NHMM, and then the experiment design. Two complementary control strategies are tested. The first uses the local Lyapunov exponent (LLE) to trigger interventions when it signals imminent eddy amplification. This provides a physics based, locally adaptive control signal. The second strategy employs a Non-homogeneous Hidden Markov Model (NHMM), which classifies latent regimes and estimates their transition probabilities conditioned on seasonal forcing. 

\subsection{\label{sec:level2}Lorenz-84 with Seasonal Forcing}
We consider the Lorenz–84 (L84) model that represents idealized mid-latitude atmospheric circulation under external forcing by the equator to pole temperature gradient and land-ocean temperature contrast. The model can be derived through a low-order truncation of the governing equations of atmospheric motion and has been widely used to illustrate key properties of atmospheric dynamics, including chaotic variability, intransitivity, and low-frequency variability. Intransitivity, in particular, refers to the existence of long-term regimes of behavior (i.e., distinct attractors with their own statistical properties) and the potential for transitions between regimes, which can lead to abrupt changes in circulation intensity and the emergence of extreme states \cite{https://doi.org/10.1111/j.1600-0870.1984.tb00230.x, James1994}. It captures the interaction between the large-scale zonal jet stream and planetary-scale eddies, which are primary characteristics of mid-latitude atmospheric circulation. Atmospheric blocking and high energy eddies are emergent phenomena in such a model, that provide a connection to weather extremes such as freezes, or heat waves, and atmospheric rivers, or frontal storms. The L84 system of equations is:
\begin{align}
\frac{dx}{dt} &= -y^{2} - z^{2} - ax + aF(t), \\
\frac{dy}{dt} &= xy - bxz - y + G, \\
\frac{dz}{dt} &= bxy + xz - z.
\end{align}

Here, $x$ represents the strength of the zonal jet stream, while $y$ and $z$ correspond to the amplitudes of the cosine and sine phases of planetary eddies. The nonlinear interaction terms ($xy, xz$) describe the amplification of eddies through energy exchange with the jet, while the quadratic damping terms ($-y^2, -z^2$) in the $x$ equation represent jet energy loss to the eddies. The terms $-bxz$ and $bxy$ capture the advection or displacement of eddies by the mean flow, with $b > 1$ implying faster displacement relative to amplification. Linear damping terms reflect mechanical and thermal dissipation, with time scaled so that the eddy damping rate is unity and the zonal flow damping rate is reduced by a factor $a < 1$. This non-autonomous formulation introduces two external forcing parameters: equator to pole temperature gradient ($F(t)$) and the land–ocean temperature contrast ($G$). The forcing could also include El Nino (ENSO) dynamics if inter-annual variations were a consideration \cite{https://doi.org/10.1111/j.1600-0870.1984.tb00230.x, Jain1998, SurfaceTemperatureGradientsasDiagnosticIndicatorsofMidlatitudeCirculationDynamics}.

To represent seasonality, we introduce time dependence into the equator to pole temperature gradient. First, we performed experiments with discrete seasonal values of F from 5 to 8, representing conditions from spring to winter \cite{Lorezn1990}. We then extend this framework by prescribing continuous seasonality with: 
\begin{align}
F(t) &= F_{0} + F_{1} \cos(\omega t)
\end{align}
where $F_0$ = 7 is the mean equator to pole temperature gradient, $F_1$ = 2 is the amplitude of the seasonal cycle, and $\omega$ is the seasonal frequency. 

To represent observational uncertainty arising from the potential divergence of uncertain initial conditions, multiplicative white noise is applied to the simulated values $\mathbf{x_t}$ from the model, with amplitude scaled to the instantaneous state magnitude $(m)$, and 
$f(\cdot)$ represents a discrete time map corresponding to the observational sampling frequency
\begin{align}
\mathbf{x}_{t} = f(\mathbf{x}_{t-1}) + \varepsilon_{t}, 
\qquad 
\varepsilon_{t} \sim \mathcal{N}\!\left(0,\, m \cdot \lvert \mathbf{x}_{t-1} \rvert \right).
\end{align}
This process explicitly embeds seasonal modulation of large-scale forcing into the L84 system while allowing noisy state information to drive regime decoding and control decisions, yielding a more realistic and stringent testbed for adaptive intervention experiments. The seasonal L84 trajectories are simulated with a fourth-order Runge--Kutta scheme with a fixed time step of $\Delta t = 0.01$, and standard parameter values $(a = 0.25,\; b = 4.0,\; G = 1.0)$. The initial state $[2.4,\; 1.0,\; 0]$ is taken from the Lorenz (1990) paper \cite{Lorezn1990}. For each experiment, we consider a total evolution of 2000 time steps from the initial condition.

\subsection{\label{sec:level2}Non-homogeneous Hidden Markov Model (NHMM)}
Because the L84 system does not remain confined to a single regime, but instead exhibits regime switching driven by its intrinsic dynamics and external forcing. To identify latent dynamical regimes in the L84 trajectory and to characterize their time-varying transition behavior, we employ a Non-homogeneous Hidden Markov Model (NHMM). The observed state vector \(\mathbf{X}_t = (x_t, y_t, z_t)\) is generated by the L84 dynamics with additive and multiplicative noise. We assume that the system evolves through \(K\) latent regimes \(S_t \in \{1,\dots,K\}\), and that both the transition probabilities and the emission distributions depend on the latent state.

Conditional on the hidden state \(S_t = k\), each L84 variable is modeled as a Gaussian AR(1) process:
\begin{align}
x_t \mid S_t = k &\sim \mathcal{N}(\mu_{x,k} + \phi_{x,k}\, x_{t-1}, \, \sigma_{x,k}^2), \\
y_t \mid S_t = k &\sim \mathcal{N}(\mu_{y,k} + \phi_{y,k}\, y_{t-1}, \, \sigma_{y,k}^2), \\
z_t \mid S_t = k &\sim \mathcal{N}(\mu_{z,k} + \phi_{z,k}\, z_{t-1}, \, \sigma_{z,k}^2).
\end{align}

The corresponding emission likelihood for state \(k\) is
\begin{equation}
b_k(\mathbf{X}_t) = 
\prod_{d\in\{x,y,z\}}
\mathcal{N}\!\left(x_{d,t};
\, \mu_{d,k} + \phi_{d,k}\,x_{d,t-1},
\, \sigma_{d,k}^2 \right).
\end{equation}

Unlike a standard Hidden Markov Model (HMM) with fixed transition probabilities,
\begin{equation}
P(S_{t+1} = j \mid S_t = i) = P_{ij}
\end{equation}

The NHMM extends this framework by allowing the transition probabilities to depend on external covariates. Let \(\mathbf{C}(t)\) be a covariate vector encoding the seasonal cycle:
\[
\mathbf{C}(t) =
\begin{bmatrix}
\sin(2\pi t/365) \\
\cos(2\pi t/365)
\end{bmatrix}.
\]

\begin{equation}
P(S_{t+1} = j \mid S_t = i,\, \mathbf{C}(t)) = P_{ij}(t).
\end{equation}

It is modeled using a multinomial logistic regression for a transition from state $i$ to state $j$ at time $t$. We define the linear predictor for the non-homogeneous transition as
\begin{equation}
\ell_{ij}(t) = \beta_{0,ij} + \beta_{1,ij}\, \mathbf{C}(t),
\end{equation}
where $\beta_{1,ij}\, \mathbf{C}(t)$ denotes the inner product between the regression coefficients and the covariate vector and the corresponding multinomial logit transition probability
\begin{equation}
P_{ij}(t) =
\frac{\exp\!\big(\ell_{ij}(t)\big)}
     {\sum_{k=1}^{K} \exp\!\big(\ell_{ik}(t)\big)}.
\end{equation}

The coefficients $\{\beta_{0,ij}, \beta_{1,ij}\}$ are extracted directly from the fitted \texttt{depmixS4} \cite{depmixS4} model, which estimates all parameters of the NHMM using the Expectation–Maximization (EM) algorithm. For the model selection, we evaluate the Penalized Likelihood of the model using the Bayesian Information Criterion (BIC), and then vary $K$ to consider different numbers of latent states, and choose the model that minimizes the BIC. The model fitting results can be found in the Appendix~\ref{secA}. 

After fitting the NHMM parameters with the EM algorithm, we classify past states and forecast future regimes employing the Viterbi algorithm using dynamic programming. The score $\delta_t(i)$ represents the maximum joint probability of any state path ending in state $i$ at time $t$. This is updated at the next step by combining the previous scores with the time-varying transition probabilities and the emission likelihoods. This procedure yields the most probable regime sequence consistent with the observed data and the seasonal covariate.

\begin{align}
\delta_t(i) &= \max_{s_1,\ldots,s_{t-1}} 
  P(s_1,\ldots,s_{t-1}, S_t=i, x_1,\ldots,x_t|\theta), \\
\delta_{t+1}(j) &= \max_i \, \delta_t(i) \, P_{ij}(t) \, b_j(x_{t+1}),
\end{align}

For forward-looking control applications, we use the forward algorithm to propagate state probabilities across a prediction window. Starting from the current distribution $\alpha_t$: 
\begin{align}
\alpha_t(j) &= P(S_t = j \mid x_{1:t},\theta)
\end{align}
The probabilities are updated recursively using the sequence of time-varying transition matrices through the specific prediction horizon ($h$)
\begin{align}
\alpha_{t+h} &= \alpha_t P(t) P(t+1) \cdots P(t+h-1),
\end{align}

The resulting distribution $\alpha_{t+h}$ provides the time-evolving probability of future regimes. This evolving probability vector is used to define \emph{dangerous} states and forms the basis of the optimal perturbation control algorithm described in Section~\ref{subsec4}.

\subsection{\label{sec:level2}Adaptive Control based on LLE triggers}

In this case, the local Lyapunov exponent (LLE) \cite{Ott2002} is a trigger for diagnosing instabilities in the L84 system. To conduct adaptive control, we follow the strategy previously described to conduct adaptive control \cite{egusphere-2025-3997} with the goal to limit the high eddy amplitude area defined by $\lvert y \rvert + \lvert z \rvert$. The key components are 
\begin{enumerate}
    \item at each time, using the current state $\mathbf{x}(t)$, estimate the LLE as a 
    measure of potential divergence;

    \item determine an upper percentile threshold of the LLE values evaluated from a long 
    experiment with the model, as a trigger for intervention or control;

    \item If control is initiated at time $t^{*}$, then solve for the timing and magnitude 
    of the perturbations $\delta \mathbf{x}(t)$ to the projected state $\mathbf{x}(t)$, 
    $t = t^{*}:t^{*}+h$, such that the total perturbation energy is minimized while ensuring 
    that the eddy-related energy remains within prescribed bounds and that the magnitude of 
    any single perturbation does not exceed a specified limit.
\end{enumerate}

Note that multiplicative noise is added for all simulations in the way previously indicated, 
including in the control stage. Here, we define four seasonal settings of the L84 system by 
varying the equator--pole temperature gradient $F$ between $5$ and $9$, corresponding to 
spring through winter \cite{Lorezn1990}. Among these, the winter configuration generates the 
most chaotic behavior, characterized by stronger eddy activity and larger amplitudes. In the 
real atmosphere, such states correspond to enhanced meridional moisture transport and can be 
associated with the development of extreme events such as hurricanes and atmospheric rivers. 
Here, the seasonal LLE framework serves as a baseline control model, against which we later 
compare the regime-based NHMM approach.

\subsection{Adaptive Control using NHMM}\label{subsec4}

The goal is to consider perturbations of the current state that limit the probability of entering undesirable regimes while minimizing the energy for perturbation, subject to constraints on the total energy for a perturbation applied at any time. At each simulation step, the current L84 state and time are passed into the NHMM-based control framework, which proceeds through three stages: danger prediction, perturbation optimization, and state update. 

After training the NHMM on a long, control-free L84 simulation, we compute for each hidden state \(i\) the empirical distribution of the eddy amplitude \(|Y|+|Z|\). We define the set of \emph{dangerous} states as \(D = \{\, i : |Y|+|Z| > E^{\ast} \,\}\), where \(E^{\ast}\) is a prescribed eddy-amplitude threshold (Appendix~A2).

For each state $i \in D$, we estimate the probability that the eddy amplitude exceeds this threshold,
\begin{equation}
w_i = P_i\!\left(|Y| + |Z| > E^{\ast}\right),
\end{equation}
which corresponds to the fraction of observations assigned to state $i$ whose eddy amplitude exceeds $E^{\ast}$.

To represent the effect of a control perturbation at time $t$, we define the perturbed state $\tilde{\mathbf{X}}_t = \mathbf{X}_t + \mathbf{u}_t$, where the control vector is $\mathbf{u}_t = (\delta x_t,\, \delta y_t,\, \delta z_t).$

The same forward recursion (Eq.~12) is then initialized from the perturbed state $\tilde{\mathbf{X}}_t$, producing the controlled predictive distribution $\alpha_{t+h}(\mathbf{u}_t)$ for each look-ahead step $h=1,\ldots,H$, where $H$ denotes the total prediction horizon.

We define the danger score at look-ahead step $h$ as
\begin{equation}
d_{t+h}(\mathbf{u}_t)
= \sum_{i\in D} w_i\, [\alpha_{t+h}(\mathbf{u}_t)]_i,
\end{equation}
where $[\alpha_{t+h}(\mathbf{u}_t)]_i$ denotes the probability assigned to state 
$i$ at time $t+h$.  
The discounted danger score at time $t$ over a prediction window of length $H$ is
\begin{equation}
Danger_t(\mathbf{u}_t)
= \sum_{h=1}^{H} \frac{1}{h}\, d_{t+h}(\mathbf{u}_t).
\end{equation}

A control action is triggered once this danger score exceeds a prescribed 
threshold $D^{\ast}$, i.e.\ $Danger_t(\mathbf{0}) > D^{\ast}$.

The control objective is to minimize the discounted multi-step danger score over a prediction window of length $H$ while penalizing the perturbation energy. This leads to the following optimization problem:
\begin{align}
\min_{\mathbf{u_t}} \quad 
& Danger_t(\mathbf{u_t}) \;+\; \lambda \,\|\mathbf{u_t}\|_2^2,
\label{eq:control-objective} \\
\text{s.t.} \quad 
& \tilde{\mathbf{X}}_t = \mathbf{X}_t + \mathbf{u_t}, \\
& \|\mathbf{u_t}\|_2 \le \varepsilon.
\end{align}

Here $\lambda > 0$ is a regularization parameter controlling the trade-off between reducing future danger and limiting the control effort, and $\varepsilon$ specifies the maximum allowable perturbation magnitude, preventing physically unrealistic adjustments to the system. The quantity $\|\mathbf{u}\|_2$ denotes the Euclidean norm of the perturbation vector.

The complete control experiment integrates all components into a sequential, time-stepping simulation. At each time step, the natural trajectory evolves according to the unmodified L84 dynamics with seasonal forcing. In parallel, for the controlled trajectory, the current state is used to propagate a distribution of future regimes over a fixed prediction horizon using the NHMM forward algorithm. If this forecast indicates an elevated probability of entering dangerous hidden states within the horizon, the control optimization problem is solved, and the resulting perturbation vector is applied to the current state. The controlled trajectory is then advanced from the perturbed state. Both natural and controlled trajectories are integrated using a fourth-order Runge–Kutta scheme. Throughout the simulation, we record whether control was activated, the magnitude and direction of the applied perturbation, and the decoded hidden state at each time.

\subsection{Experiment Setup}\label{subsec5}
Based on the introduction above, we advance several methodological innovations within the Lorenz-84 (L84) framework to explore adaptive chaos control under non-autonomous, seasonally forced conditions. The experiment procedure as follows:
\begin{enumerate}
    \item First, we introduce seasonal variability in the equator to pole temperature gradients that provide the forcing to the model. This is important to address the path dependence \cite{Jain1998} in the model trajectories and state variable statistics introduced by the non-autonomous forcing of the model. 
    \item Second, recognizing that the model represents a deterministic kernel of the actual dynamics, we introduce noise with amplitude proportional to the state variable magnitude that the control algorithm needs to contend with. 
    \item Third, we consider the deterministic divergence characteristics of the system through the LLE, and the stochastic transition characteristics through the non-homogeneous Hidden Markov Model (NHMM), and consider the use of either as a criteria for exercising model control and explore their complementarity in this non-autonomous context. The seasonal cycle is the covariate for the NHMM, allowing the identification of latent states over the year whose transition probabilities change seasonally. 
    \item Fourth, we consider that the goal of adaptive chaos control can be 1) avoidance of a transition to an undesirable regime, e.g., one that may have adverse consequences, or 2) limiting the total energy associated with the eddies in the L84 model, that may conceptually represent powerful tropical or mid-latitude storms coupling with the jet stream in the model. 
    \item Finally, we adopt a two stage finite time control process where in the first stage, the control is triggered by a LLE threshold being crossed based on an analysis of the model integration over a future finite time from the current state, or if the indicated transition probability to a hidden state of concern for the NHMM. The second stage then solves for a perturbation schedule such that the energy associated with each perturbation is bounded, the total perturbation energy over the horizon is minimized and constraints are applied to bound the future states over the next time period. Noise is added to the trajectory at every time step and the control strategy is re-applied sequentially at every time step. 
\end{enumerate}

Specifically, our control objective is not only to suppress excessive eddy growth through perturbations but also to identify and anticipate transitions into dangerous states, bringing the experiment closer to the challenge of controlling weather extremes in an idealized setting. In the physical atmosphere, such perturbations could be implemented via latent-heat release via cloud seeding or other methods that induce differential local temperature gradients. Recent studies have demonstrated that latent heating is a critical driver of strengthening the subtropical jet and modulating Hadley-cell variability \cite{Lin2025, https://doi.org/10.1029/2025GL116437}.

\section{Results}\label{sec3}
\subsection{LLE-based Control Result}\label{subsec6}
\begin{figure}[tbp]
  \centering
  \includegraphics[width=\columnwidth]{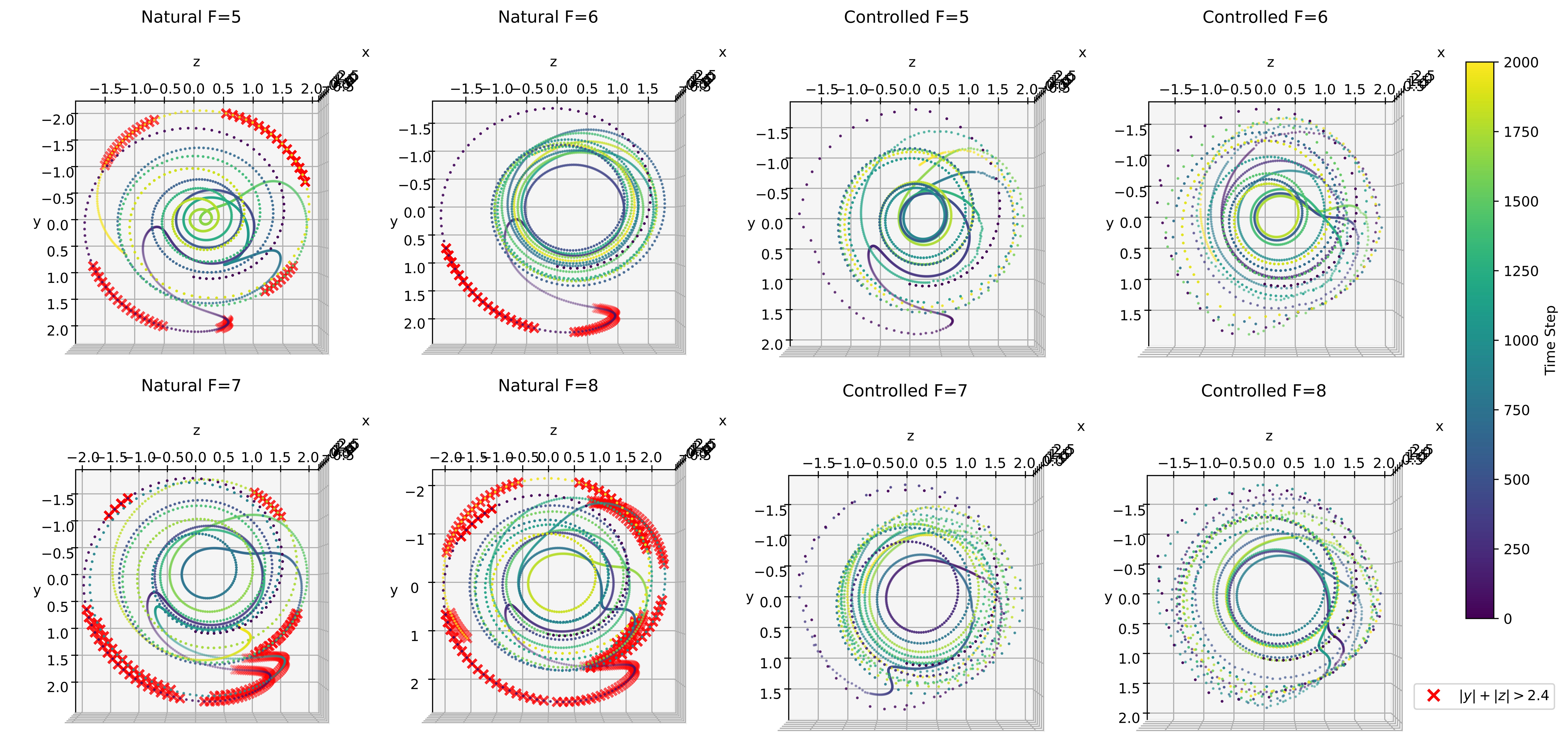} 
  \caption{L84 trajectories under natural dynamics (left) and LLE-based control
  (right) across seasonal forcing values $F=5,6,7,8$. Colors indicate time
  progression, with red crosses marking the 90th percentile eddy amplitude threshold.}
  \label{fig:l84-trajectories}
\end{figure}

In Fig 1, we illustrate the L84 trajectory under four different forcings representing four seasons. In the natural runs, larger forcing values (winter-like conditions) produce stronger and more frequent excursions into high eddy amplitude regimes, as indicated by repeated threshold crossings. To characterize such extremes, we define the threshold as the 90th percentile of $|Y|+|Z|$, which in our simulations corresponds to an LLE of approximately $2.4$. Under LLE-based control, trajectories remain confined within the threshold of high eddy regime, which indicates effective suppression of dangerous eddy growth across all seasonal backgrounds. The color progression highlights how control modifies the temporal evolution, reducing transitions into extreme states while preserving the intrinsic oscillatory variability of the system. These results show that LLE-based control can limit the amplification of instabilities into extreme eddy events, providing evidence that small, targeted perturbations can robustly regulate the L84 system across a range of seasonal forcing values.

\subsection{NHMM based Control Result}\label{subsec7}

\begin{figure}[tbp]
  \centering
  \includegraphics[width=\columnwidth]{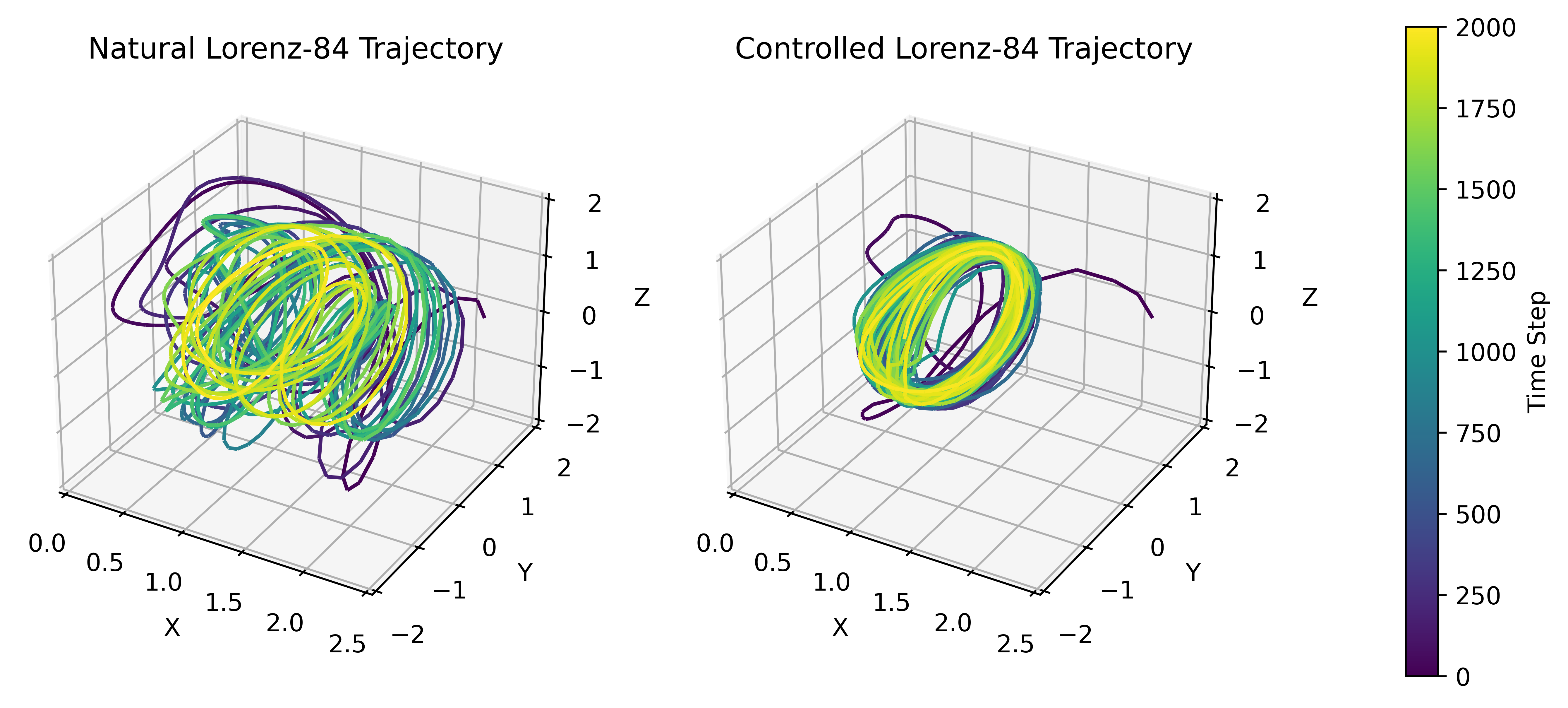} 
  \caption{L84 trajectories under natural dynamics and NHMM-based control (Colors indicate time progression); red markers denote time steps where control was applied.}
  \label{fig:nhmm_result}
\end{figure}

We examine the effect of NHMM-based control on the L84 dynamics. The uncontrolled trajectory (left panel of Fig 2) explores a broad portion of phase space, with frequent irregular excursions, which reflect the system’s intrinsic instability under seasonal forcing. With NHMM-based control, the trajectory reorganizes into a more confined structure over the time steps. Instead of suppressing all variability, control is applied selectively at time steps where the predicted state distribution indicates high likelihood of entering dangerous hidden regimes. This outcome demonstrates the regime-aware advantage of NHMM control: by anticipating transitions based on hidden-state dynamics, the method reduces the frequency and intensity of extreme excursions while maintaining realistic variability.

\subsection{Control Result Dashboard}\label{subsec8}
\begin{figure}[tbp]
  \centering
  \includegraphics[width=\columnwidth]{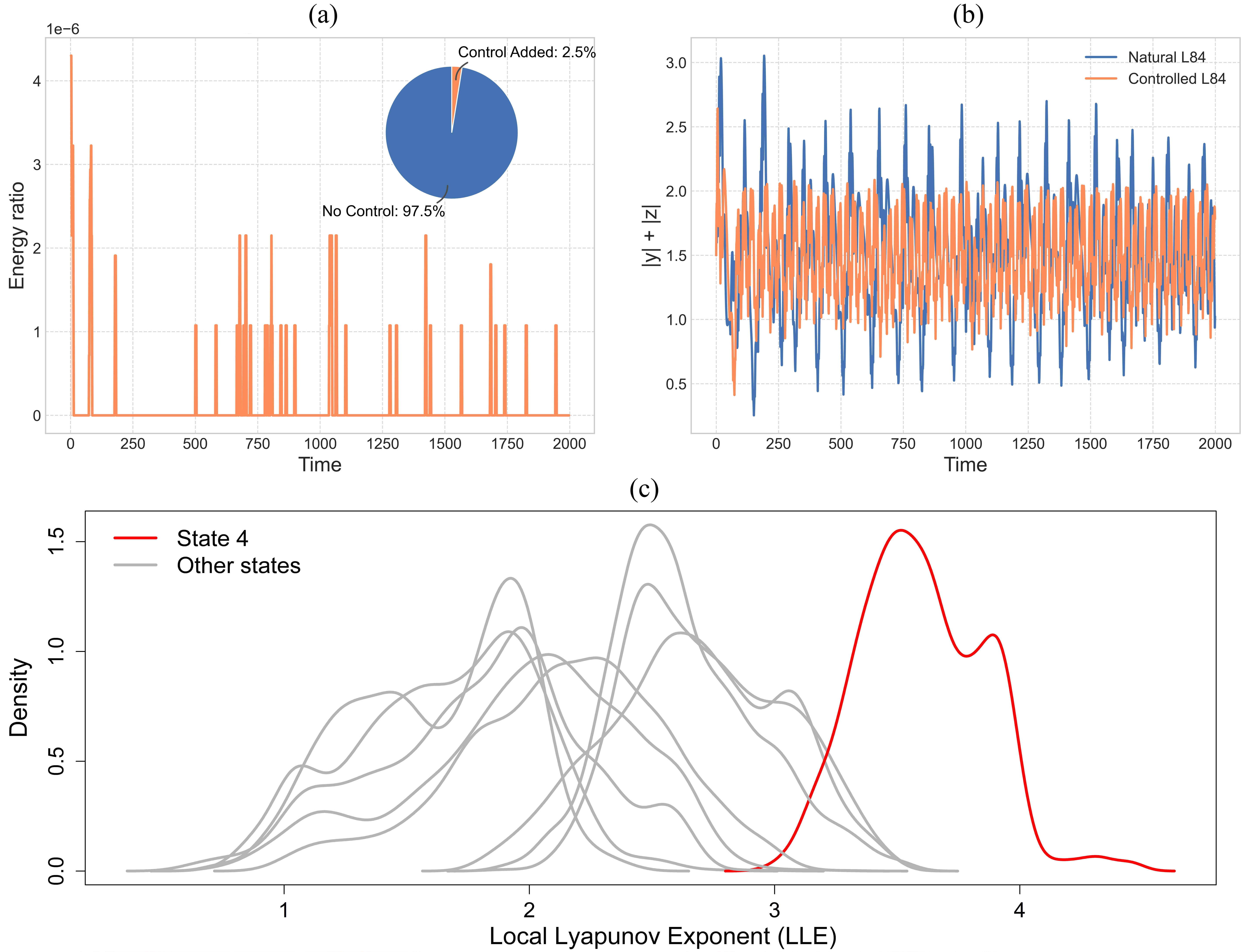} 
  \caption{NHMM-based state analysis and control performance: Time series of the energy ratio diagnostic (a); Eddy amplitude for natural and controlled trajectories (b); Kernel density estimates of LLE conditioned on NHMM states (c).}
  \label{fig:nhmm_Dashboard}
\end{figure}
The energy-ratio diagnostic (Fig 3a) confirms the design goal of applying minimal control effort: interventions account for only 2.5\% of simulation steps, with control energy negligible relative to the system’s total energy. Under natural dynamics (Fig 3b), the eddy amplitude exceeds the high threshold value of 2.4 from time to time, indicating recurrent excursions into unstable regimes. With NHMM-based control, these extreme peaks are substantially reduced, with amplitudes consistently suppressed below the natural extremes. This demonstrates that the control system effectively redirects trajectories into safer regimes through small, targeted perturbations.

A key question is whether the NHMM’s hidden regimes correspond to identifiable dynamical chaotic structures of the L84 system. To address this, we compare the distribution of LLEs across decoded NHMM states (Fig 3c). While most states exhibit LLEs distributions centered between 1.5 and 2.5, state 4 is distinct with a sharp density peak above 3.5. This separation indicates that state 4 captures the system’s most unstable regime, dominated by rapid divergence of trajectories and intense eddy growth, consistent with its classification as a “danger” state in the NHMM analysis.

Taken together, these results demonstrate strong convergence between the two approaches. The LLE provides a local, physics-based measure of instability, while the NHMM offers a regime-based probabilistic framework that incorporates temporal context and covariate dependence. This confirms that NHMM states are dynamically interpretable and that both LLE and NHMM can serve as robust, complementary triggers for targeted control.

\section{Discussion and Conclusions}\label{sec4}
Our results demonstrate that NHMMs provide a useful complement to LLE-based diagnostics for adaptive control of a chaotic non-autonomous model, accounting for the observations and probability distributions of the latent regimes. The robustness of the result shows that NHMMs can work well under both seasonal forcing and multiplicative noise. This study highlights the potential for extending such approaches to more sophisticated machine learning models, in which hidden states can explicitly encode regime dynamics and inform predictive control strategies. 

Potentially, this framework can be extended from low-order dynamical systems to real climate and weather models, including state-of-the-art foundational model architectures. We introduce our new hypothesis, Weather Jiu-Jitsu, which aims to subtly redirect or diffuse hazardous atmospheric trajectories through small, strategically timed perturbations \cite{huang2025weatherjiujitsuclimateadaptation}. The underlying principles are grounded in chaos control theory. In foundational weather models such as Aurora, hidden states can be fetched from effective transformer architecture \cite{Bodnar2025, SwinTransformer}. The NHMM logic presented here could therefore be integrated with these latent representations to identify and anticipate “dangerous” regimes within high-dimensional spaces. In this way, the Weather Jiu-Jitsu provides a pathway for translating insights from idealized toy models to operationally relevant frameworks for adaptive weather and climate control.

As the results indicate, the required control energy is minimal, yet the outcome produces a substantial difference in system behavior. In physical terms, the perturbations considered could be induced by latent-heat modifications akin to targeted cloud-microphysical interventions. Recent studies show that latent-heat release actively shapes large-scale circulation by strengthening subtropical jets and destabilizing Hadley-cell structure \cite{Lin2025, https://doi.org/10.1029/2025GL116437}. Together, these findings suggest a plausible physical pathway through which small, strategically applied perturbations could be used to influence or divert extreme weather events within real atmospheric regimes.

At the same time, several limitations must be acknowledged. The L84 model is highly simplified and cannot capture the complexity of atmospheric circulation. Our definition of “dangerous” is heuristic, based on thresholds in eddy amplitude and state classification. Moreover, real-world interventions, whether through cloud seeding, laser induced heating, boundary layer modification, or pressure perturbations, involve physical mechanisms that are not represented in the present framework. Implementing an optimization algorithm to solve for the time, location and magnitude of perturbation in a full dynamical model of the atmosphere is computationally not practical, and hence we need alternate approaches to assess the feasibility of Weather Jiu-Jitsu. 

A number of weather forecasting models have been trained on extensive simulations of physics based models and are outperforming the physics based models that they emulate, especially as the forecast lead time increases (e.g. Aurora, AIFS, Earth AI) \cite{Bodnar2025, lang2024aifsecmwfsdatadriven, bell2025earthaiunlockinggeospatial}. These models permit relatively rapid computation and provide access to a number of the key variables associated with the circulation in a 3-dimensional space-time setting. The models are parameterized via a space-time latent space embedding that is exploited in somewhat different ways across implementations. The transformer architecture provides access to the underlying latent states, and to a deterministic forecast given a global initial condition. Our experiments with this architecture reveal a rather promising forecast capability, and we are exploring how noise and ensemble forecasting could be efficiently integrated. We expect that estimated Finite Time/Space Lyapunov Exponents and the latent state evolution probabilities (with noise considered) will collectively enable the development of at least a heuristic structure that can indicate promising perturbation and control strategies, going beyond the toy models explored to date. 


\appendix
\setcounter{figure}{0}

\section{NHMM model selection}\label{secA}
\subsection{NHMM model selection}\label{subsec9}
We estimated non-homogeneous hidden Markov models with different numbers of hidden states and selected the 9-state specification based on the Bayesian Information Criterion (BIC). Fig A1 shows the state-conditioned distributions in the $(y,z)$ plane, where color shading denotes the local point density within each state.
\begin{figure}[tbp]
  \centering
  \includegraphics[width=\columnwidth]{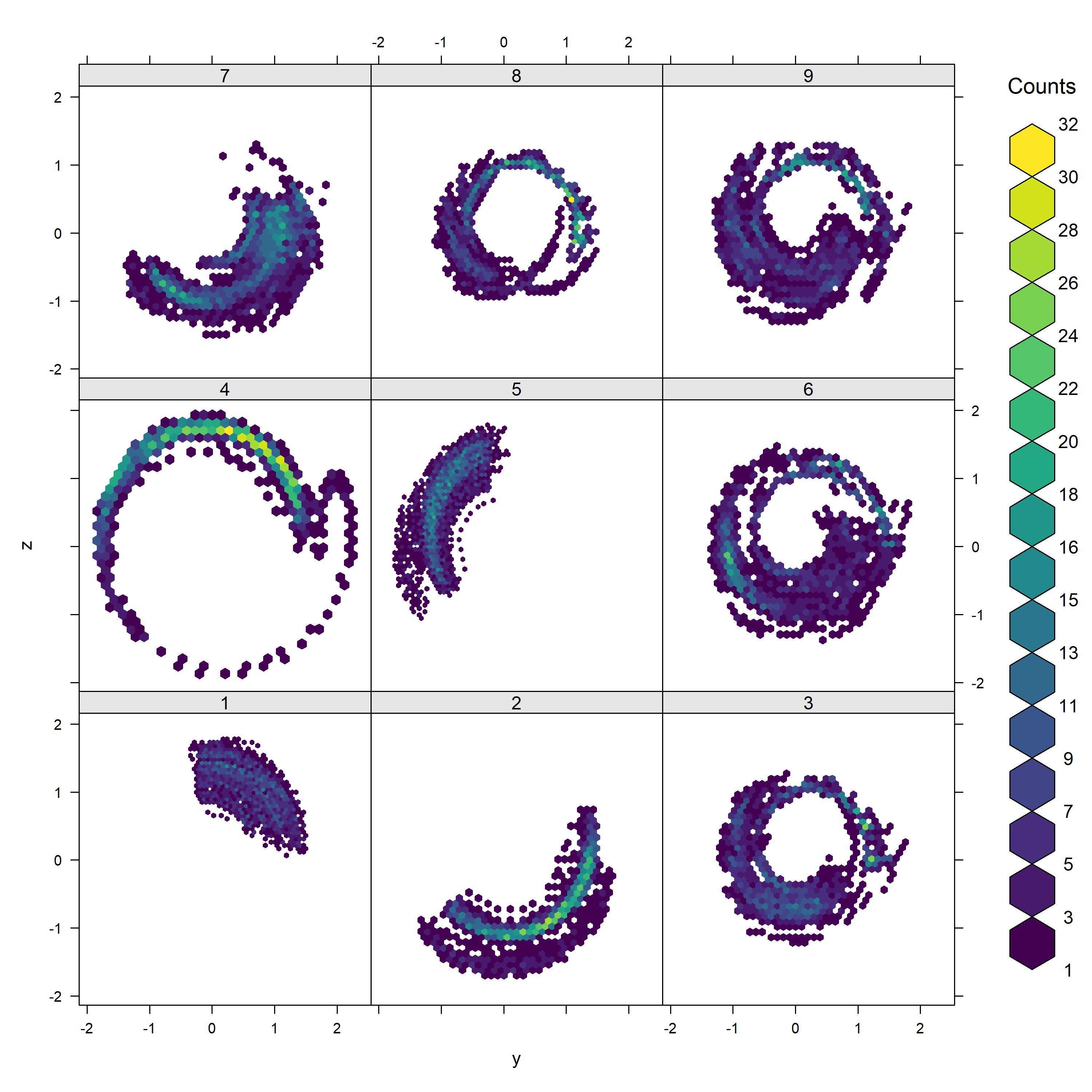}
  \caption{Spatial distribution of all hidden states, with color indicating the spatial density of each state}
  \label{fig:state_hexbin}
\end{figure}

\subsection{Dangerous State Identification}\label{subsec10}
We generated 20000 data points and defined the severity threshold using the 97th percentile $|Y|+|Z|$ as the high eddy amplitude regimes. All points above this threshold were extracted and assigned to their corresponding NHMM states showing at Fig A2. Among these exceedances, state 4 accounts for more than 90.5\% of all events above the threshold. Combined with its spatial structure in Fig A1, we designate state 4 as the “dangerous” state for subsequent control experiments.
\begin{figure}[tbp]
  \centering
  \includegraphics[width=\columnwidth]{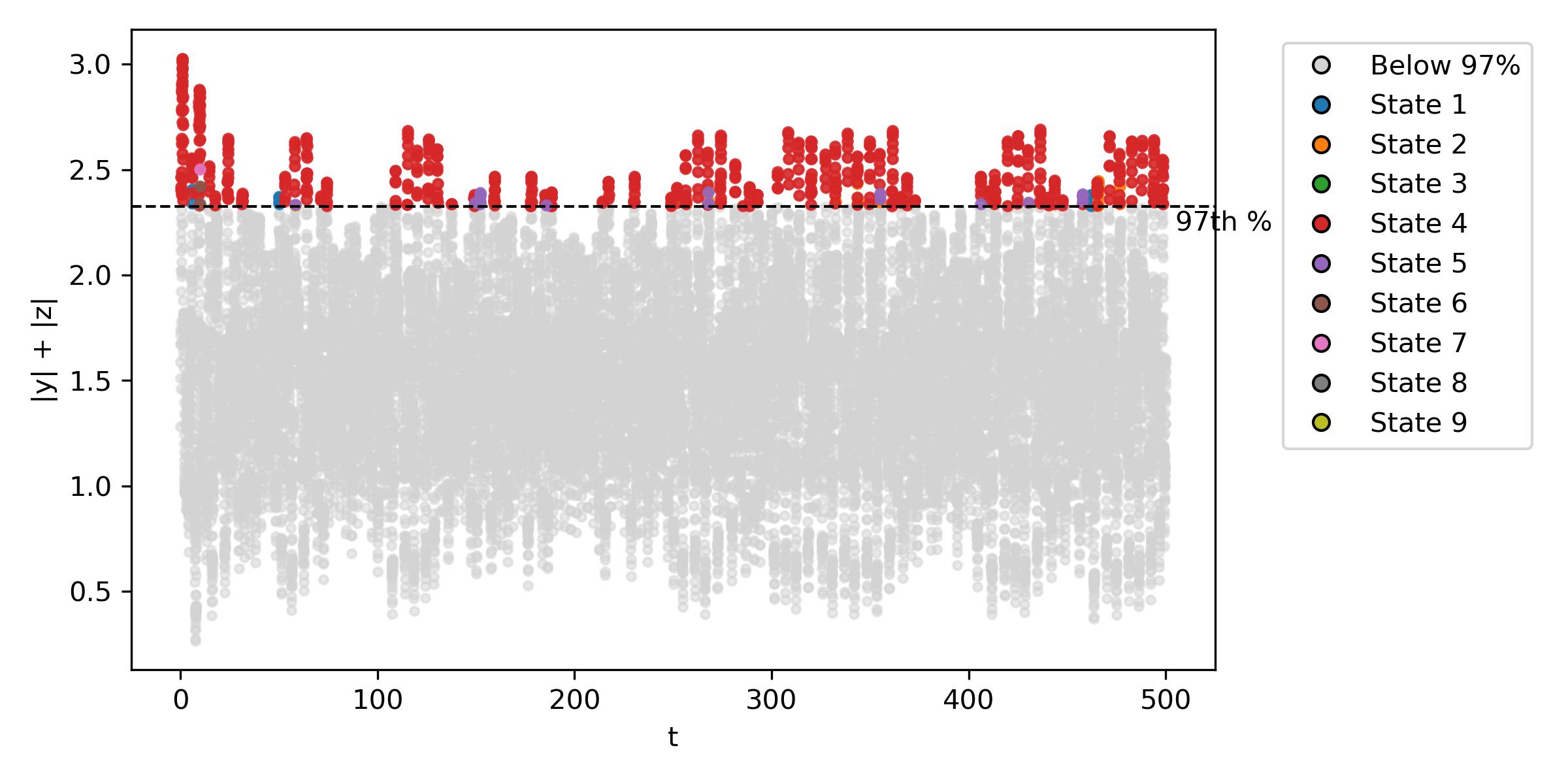}
  \caption{Time series of eddy amplitude with a dashed line marking the threshold that defines the danger regime}
  \label{fig:eddy_energy_states}
\end{figure}
\clearpage
\bibliography{prlreference.bib}

\end{document}